\documentclass{article}

\usepackage{PRIMEarxiv}

\usepackage[utf8]{inputenc} 
\usepackage[T1]{fontenc}    
\usepackage{hyperref}       
\usepackage{url}            
\usepackage{booktabs}       
\usepackage{amsfonts}       
\usepackage{nicefrac}       
\usepackage{microtype}      
\usepackage{lipsum}
\usepackage{fancyhdr}       
\usepackage{graphicx}       
\graphicspath{{media/}}     

\title{Application of  Super-Sampling to Microscopy Images Produces Image Resolution below Optical Diffraction Limit
\thanks{\textit{\underline{Citation}}: 
\textbf{Authors. Title. Pages.... DOI:000000/11111.}} 
}

\author{
  James N. Caron \\
  Quarktet \\
  Silver Spring, Maryland, USA\\
  \texttt{Caron@Quarktet.com} \\
}

\begin{document}


\title{Application of  Super-Sampling to Microscopy Images Produces Image Resolution below Optical Diffraction Limit} 


\maketitle 

\begin{abstract}  
Image Phase Alignment Super-Sampling (ImPASS) is a computational imaging algorithm for converting a sequence of displaced low-resolution images into a single high-resolution image.  The method consists of a unique combination of Phase Correlation image registration and SeDDaRA blind deconvolution.  The method has previously been validated in simulations and applied successfully to images captured in a laboratory setting. As discussed here, the performance of ImPASS surpasses similar methods that provide quantitative results.   ImPASS is applied for the first time to images taken by a widefield microscope, requiring no customization other than a translation stage, to determine if this approach can subceed the diffraction limit for this application. The 80-frame image sets had as targets a slide with a slice of Porcine Cornea, and a standard US Air Force resolution chart, providing quantitative and quantitative assessments.  The sets were up-sampled by a factor of eight, aligned, combined, and processed.  The measurement revealed that image resolution improved by a factor of 2.68 and subceeded the diffraction limit by a factor of 1.79.
\end{abstract}

\section{Introduction}

The resolution of a digital optical imaging system is limited by both the digital sampling rate of the sensor bounded by the size of a pixel, and the diffraction limit as defined by the front aperture diameter.  Producing image resolution beyond the capabilities of the imaging system has been an active research area for decades~\cite{schuler,patan,huang,cheeseman} with applications ranging from microscopes~\cite{microscopes} to space-born telescopes~\cite{telescopes}.  In recent papers, we have demonstrated a computational imaging method designated Image Phase Aligned Super-sampling (ImPASS)~\cite{ao_2020,pending} that has potential benefits for both telescopes and microscopes.  In contrast to existing super-resolution microscope (SRM) techniques, this approach does not depend on active illumination of the sample,~\cite{aip_2022} allowing the user to specify the source, which could be especially useful for multispectral imaging.  The goal of this effort is to demonstrate that ImPASS and similar methods can be considered as alternatives to some SRM methods.  

\begin{figure}[htbp]
\center{\includegraphics[width=26pc]{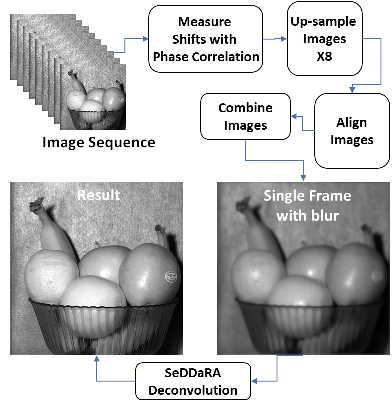}}
\caption{ Flow chart for ImPASS algorithm.  A series of displaced low-resolution images are up-sampled, aligned, and combined to form a seemingly blurry single image. Application of SeDDaRA deconvolution produces a high-resolution result. (Image Source: Author)}\label{process}
\end{figure}

In this paper, for the first time, ImPASS is applied to image sequences captured with a microscope to demonstrate the qualitative effectiveness and quantitative image resolution improvement.  A set of displaced images of a 5 micron-cut slice of a porcine cornea was taken with a 4X magnification lens with the processed result compared qualitatively to a truth image taken with a 10X lens. A second set of images was taken of a USAF resolution target to quantify the resolution improvement and compare the value to the optical diffraction limit.

The definitions for `super-resolution' is often inconsistent in the literature.~\cite{driggers, schuler}  In 1993, Sheppard et al.~\cite{shep1} defined super-resolution as ``the removal of blur caused by a diffraction-limited optical system along with meaningful recovery of the object's spatial-frequency components outside the optical system's passband.''  Many articles that followed referred to methods where low-resolution images are combined computationally producing a single high-resolution image as `super-resolution' without regard to the diffraction limit.~\cite{borman, sina1, shah}  For microscopy, however, super-resolution has been defined as achieving resolution  below the system's optical diffraction limit~\cite{bo,hell}  In 2005, Driggers et al.~\cite{driggers} attempted to remedy this conflict by referring to the computational method as `geometric super-resolution' and optical methods as `optical super-resolution'.  Unfortunately, this language was not adopted by the community as evident in later papers.   For further confusion, the term `super-resolution' has also been adopted by the gaming industry for methods that create computer-generated details to enhance visual aspects of gaming.   To be clear, particulary since the focus of this article is on a computational method to microscopy, we adhere to the following succinct definitions. 
\begin{description}
  \item[Super-resolution] A computational imaging or optical technique that produces an image with resolution beyond the diffraction limit of the optical system.
  \item[Super-sampling] A computational imaging technique that combines a sequence of images to produce a single larger image with resolution beyond the digital sampling frequency of the sensor.
\end{description}
This follows convention set by Schuler~\cite{schuler} and followed by Li.~\cite{li}

\section{Background}
The essential elements of the process are depicted in Figure~\ref{process}.  The input is an image sequence where frames have non-integer displacements produced when the camera moves relative to the scene, or in this case the scene moves relative to the camera. Accurate displacements are needed to achieve sub-pixel accuracy.  These values can be provided by the instrument if sufficiently accurate, but backlash and jitter can produce errors.  Alternatively, the displacements can be determined using image registration methods.~\cite{medha} ImPASS employs the phase correlation method~\cite{reddy, tong} to measure frame-to-frame displacements as this Fourier-based algorithm is comparatively efficient and achieves sub-pixel accuracy. 

The sequence is up-sampled to a higher sampling size using common image interpolation methods, aligned, and then combined into a single image.  The current iteration of ImPASS using cubic convolution for interpolation with the cubic parameter set to -0.5, based on the recommendation by Park and Schowengerdt.~\cite{park}  However, variations of this parameter on the order of $\pm 0.1$ do not significantly affect results.  The image combination for this study was achieved by averaging the up-sampled aligned frames.

The blind deconvolution method, SeDDaRA (Self-Deconvolving Data Restoration Algorithm)~\cite{ao_2002} is applied to the combined image to reveal the sub-pixel resolution.  SeDDaRA is a non-iterative deconvolution method that extracts a point spread function from a blurred image by comparing the spatial frequency distribution of the blurred image with that of a `reference' image. The reference image need not appear similar to the image, but only possess the appropriate distribution.  The important aspect for this application is that the reference image can have higher spatial frequencies than the target image allowing these frequencies to be recovered.~\cite{ao_2020}  

When applied to a simulated sequence of images,~\cite{ao_2020} ImPASS produced resolution close to 1/16th of a pixel, well beyond the digital sampling frequency.  When applied to empirical images with broadband illumination,~\cite{aip_2022} ImPASS produced image resolution that was 1/3rd of the original image resolution, and subceeded the diffraction limit by a factor of 1.6 with the aperture set to an $f/12$.  With a 70-nm bandwidth filter placed in front of the lens, ImPASS produced resolution that was 1/3.2 times the original filtered image and subceeding the diffraction limit by a factor of 2.63 ($f/6$).  Reference~\cite{ao_2020} provides insight on how ImPASS reconstructs the higher resolution information.

An open question from previous work is whether ImPASS can be applied to microscopy.  Compared to a camera lens, a microscope has considerably more optical elements, smaller focal length, higher magnification, and smaller aperture. Optical aberrations, such as spherical, coma, barrel distortion, and chromatic occur at different ratios in camera lenses and microscope objectives, and could be a limiting factor for ImPASS.  These differences provide sufficient reason to question, as some have, whether ImPASS, having achieved super-resolution with a machine vision camera, can produce similar results with widefield microscopy.  

\section{Methods of Super-Resolution Microscopy}
Research and application of super-resolution microscopy (SRM) has increased greatly since inventors of the technique received a Nobel prize in~2014.~\cite{microscopes}  The most well-known methods are stimulated emission depletion (STED) and single-molecule localization microscopy (SMLM),~\cite{sheng} which use lasers to generate fluorescence in samples.  Since a full review is beyond the scope of this paper, the reader is directed to articles by Valli et al.~\cite{jessica} and Prakash et al.~\cite{prakash} for a thorough description of laser-based SRM techniques and applications.  These methods are very effective but do require lasers with intricate optical setups where costs can restrict their use. Their functionality also depends on the application where, for example, direct laser light can cause photo-bleaching in the sample.~\cite{kirti} 

Super-sampling fits into the class of ensemble techniques~\cite{jessica} similar in some respects to Fourier Ptychography (FP)~\cite{konda} and Structured Illumination Microscopy (SIM).~\cite{joby} These methods also convert sequences of low-resolution images into a single high-resolution image, but do so by changing the illumination on the sample per frame. FP uses an LED array to illuminate the specimen at different angles~\cite{konda} whereas SIM illuminates the specimen with a varying moiré-like sinusoidal pattern.  Qui et al.~\cite{qiu} presented Translation Microscopy that combined elements of super-sampling with fluorescent imaging.  

The above techniques require some form of active illumination to achieve resolution improvement.  In contrast, super-sampling, being dependent on displacements, is not dependent on the illumination source.  The sample can be illuminated with the light source or bandwidth that brings out the desired feature of the sample.  As such, this technique is uniquely suited for a wider variety of applications including multi-spectral and non-visible imaging and in combination of methods such as optical coherence tomography and confocal microscopy.

\section{Comparison to Similar Computational Methods}   
Super-sampling methods generally differ in the methods of alignment and deconvolution.~\cite{kamal} Methods of validation of the approaches also vary.  In the 1990s and early 2000s, validation was demonstrated by visual inspection of the images, before and after processing.~\cite{schuler, cheeseman, shep1, sina1, jiang} While achieving the goal of validating the result, this relied on subjective observation.  Around 2005, researchers started to use similarity measures to demonstrate improvements.~\cite{baba, hehu, protter, gao, ngoc}  A survey by Nasrollahi and Moeslund~\cite{kamal} provides a list of these metrics, most notably Mean Square Error, Peak-Signal-To-Noise, and structural similarity index measure (SSIM).  These metrics are useful in evaluating an algorithm when a truth image is available.~\cite{kohler}  Nasrollahi states that while these two measures have been used often by researchers, the metrics do not represent the human visual system  adequately~\cite{kamal}. As such, higher similarity scores do not directly correlate to resolution or resolution improvement.~\cite{woods}

In 2014, Carles et al~\cite{carles} applied a super-sampling method consisting of SIFT alignment and Maximum Likelihood deconvolution to images from a 5~by~5 camera array with $f/2.8$.  Resolution improvement was measured using the Michelson method where a group of bars in a resolution chart are deemed resolved if the contrast value is below 0.81.   Contrast is calculated  with 
\begin{equation}\label{eq5}
  C = \frac{L_{MAX}-L_{MIN}}{L_{MAX}+L_{MIN}}
\end{equation}
where $L_{MAX}$ is maximum luminance, and $L_{MIN}$ is minimum luminance of a bar group.  By comparing groups that meet the resolution criteria, the group measured 2.25 times improvement where achieving the diffraction limit required a 4.5 factor of improvement.  

In 2020, Stankevich~\cite{stan} applied a multi-step super-sampling process to a sequence of satellite images reporting a 1.49 factor of improvement in resolution.  To arrive at this value, the researchers took an edge spread function directly from their images and assessed the related modulation function.  The diffraction limit of the system is not discussed.  In 2025, Brewer et al~\cite{brewer} used a variation of Shift-and-Add with Phase Correlation  to report a 1.42 improvement in resolution.  Resolution improvement was determined using valley-to-peak ratio of bars on a USAF chart. 

In our 2022 paper,~\cite{aip_2022} we measured image resolution using the slant-degree method described by ISO 12233.~\cite{iso}  This approach allowed both qualitative comparisons between original and processed images, and quantitative comparison to the system's diffraction limit.  Measurements were taken as a function of $f/\#$ ranging from 2.9 to 22.8.  For broadband illumination, the resolution improvement ranged from 3.23 to 1.23 with an average improvement of 1.95.  For narrow-band illumination, improvement ranged from 3.24 to 1.43.  This exceeds the results from Stankevich and Brewer.  Whereas Carles found an improvement of 2.24 using broadband resolution and an $f/\#$ of 2.8, we found an improvement of 3.24 for $f/2.9$.    

\begin{figure}[htbp]
\centering \includegraphics[width=26pc]{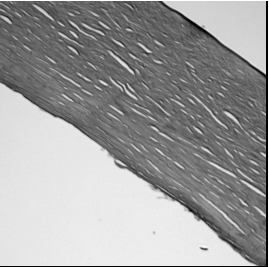}
\caption{ The microscope slide contains a 5 micron-cut slice of a porcine cornea illuminated by broadband visible light.  This image is a single frame of the sequence, cropped to $532^2$ pixels.  }\label{control}
\end{figure}              
   
\section{Application to Porcine Cornea Image Sequence}
ImPASS was applied to a sequence of translated images of a microscope slide featuring a 5~micron-cut slice of a porcine cornea originally intended for histology and thus stained with hematoxylin and eosin.  The thickness of the cornea is about 800~microns, mostly consisting of collagen fibers and lamellae. In the anterior third  collagen, lamellae are interwoven in different directions, while in the posterior two thirds, collagen lamellae are mostly arranged parallel to the cornea surface.

The Olympus Inverted Fluorescence Microscope IX-83 had an Olympus UPlanFL 4x lens with a numerical aperture of 0.13.  The sample was illuminated with broadband visible light and placed on an Olympus IX3-SSU translation stage.  Eighty-one images were taken with the Hammatsu ORCA-Flash 4.0 v3 camera while the sample was translated in diagonal steps of 989.95~nm corresponding along 2 diagonals.  Pixel size on the focal plane array was 6.5~$\mu$m by 6.5~$\mu$m. Each image  has an original frame size of 2034 X 2042 pixels. The intention is to up-sample the images by a factor of 8 which would produce an 81-frame sequence of 16262 X 16336 pixel-sized images. That is a bit taxing for a desktop computer, so images were cropped to $532^2$ pixels (shorthand for 532 X 532 pixel area) for easier processing.  The cropped image with original resolution is shown in Figure~\ref{control}.

While the algorithm can be automated with a single program, three software programs were utilized for this study.  The image registration was conducted with a script using Digital Optics V++.~\cite{vpp} L3Harris IDL~\cite{l3harris} was employed to interpolate and combine images as well as make final resolution measurements.  The deconvolution step was conducted using Quarktet Tria.~\cite{tria}

\begin{figure}[htbp]
\center{\includegraphics[width=34pc]{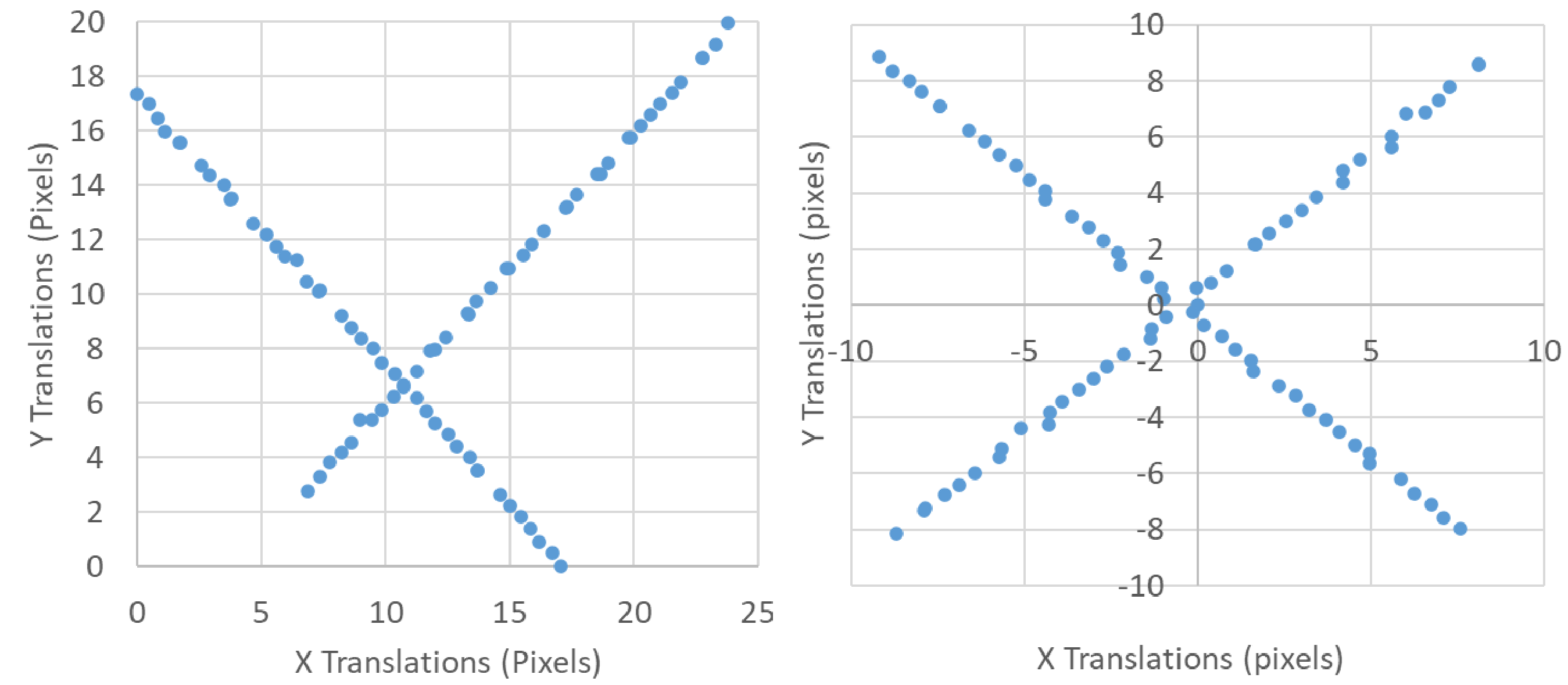}} 
\caption{ (Left) The relative positions of the translated porcine cornea images as measured using phase correlation, showing some backlash.    (Right) The relative frame positions for the set of resolution chart images. The positions are displayed with unit of pixels to demonstrate effective pixel sampling in the image plane.  }\label{regress}
\end{figure}

Displacements between frames are required to align the images to a single reference frame.  Two methods of registration were attempted.  In the first, the sequence was processed using the output from the translation stage controller.  While this output had equally-spaced displacements, it was apparent from visual inspection that the actual image displacements were not always in step with the output.  The second attempt used phase correlation to determine the relative displacements between the frames.  There were no significant differences in the results suggesting that the level of error between the two approaches were similar.  The results presented here used the translations from phase correlation.  The graph in Figure~\ref{regress}(left) shows the relative positions of the frames with average translations of 0.428 steps per pixel.

\begin{figure}[htbp]
 \centering \includegraphics[width=26pc]{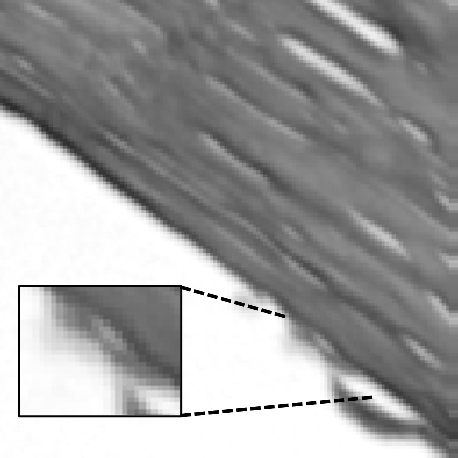}\\ \centering \includegraphics[width=26pc]{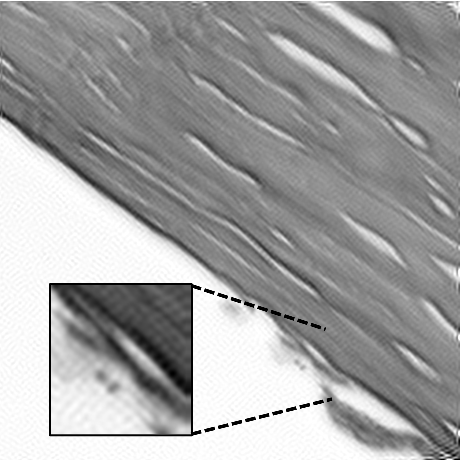}
\caption{ Comparison of the original resolution to the processed resolution.  The top image is a $128^2$ pixel area of a single unprocessed frame.  The bottom image is the same region but after processing.  The $1024^2$ pixel area shows features that cannot be discerned at the original resolution.}\label{result}
\end{figure}

Phase correlation is a Fourier-based registration method.~\cite{reddy} Given two images, $I_c(m,n)$ and $I_t(m,n)$ with non-integer translational difference of $(\Delta x, \Delta y)$, the images are related by
\begin{equation}\label{feq1}
    I_t(m,n)=I_c(m-\Delta x,n-\Delta y).
\end{equation}
A Fourier transform is applied to both sides of the equation produces
\begin{equation}\label{feq2}
    F_t(u,v)=e^{-i(u \Delta x+ v \Delta y)}F_c(u,v)
\end{equation}
where $(u,v)$ are the coordinates in frequency space. The translation differences appear as phase changes in frequency space and  can be derived from the phase
difference between the transformed images. The phase difference is
equivalent to
\begin{equation}\label{feq2a}
    e^{(u \Delta x+ v \Delta y)}=\frac{F_c(u,v)F^*_t(u,v)}{|F_c(u,v)F^*_t(u,v)|}
\end{equation}
where $*$ represents the complex conjugate.  This difference between the phase components appears as a sinusoidal pattern.  Application of another Fourier transform
to the phase difference produces a correlation peak whose coordinates are $(\Delta x, \Delta y)$ . Application of a centroid function to the correlation peak provides the sub-pixel location.

\begin{figure}[htbp]
\centering \includegraphics[width=26pc]{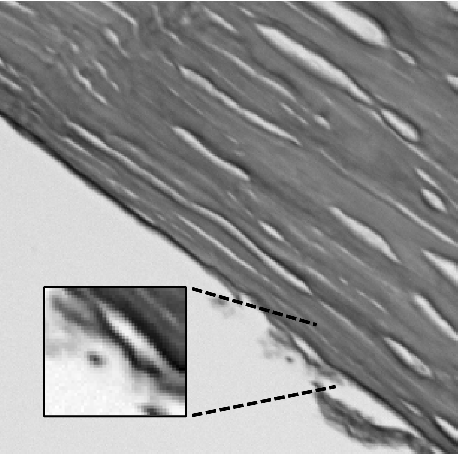}
\caption{A truth image taken with a 10X lens cropped to the same region as Figure~\ref{result}. The subframe shows a feature that is resolved in the processed image, but not in the 10X truth image.}\label{lens10}
\end{figure}

The use of phase correlation does not preclude the use of other image registration methods, but phase correlation has proven to be robust, efficient, and achieves sub-pixel accuracy.~\cite{tong}  A full survey of registration methods and their effectiveness and utility for super-sampling is beyond the scope of this study.

Processing of the 81 images took about 40 minutes on a standard desktop PC with up-sampling being the most computer-intensive step.  Each frame is up-sampled by a factor of $M_{SS}=8$ and aligned.  SeDDaRA was applied to the combined image using the parameters RofI (radius of influence) set to 24 pixels and $C_2$ set to 0.02.  These parameters are set to limit artifact and noise amplification, and are determined through trial and error.  Figure~\ref{result}(bottom) shows a cropped $1024^2$ pixel portion of the result alongside an $128^2$ pixel central portion of an unprocessed frame (top).  The result is clearly super-sampled showing many features that are not apparent in the original image.  A truth image taken with the microscope's 10X lens is shown in Figure~\ref{lens10}.  The resolution of processed result is comparable to that of the 10X image.  In fact, there is a feature, shown in the insets that appears as two dots in Figure~\ref{result}(bottom), but are not resolved in Figure~\ref{lens10}. This suggests that the processing has improved the resolution by at least a factor of 2.5.

\begin{figure}[htbp]
\centering \includegraphics[width=24pc]{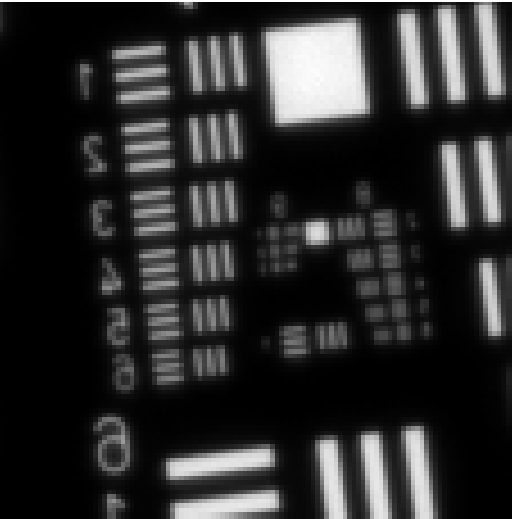} \\ \vspace{1pc} \centering \includegraphics[width=24pc]{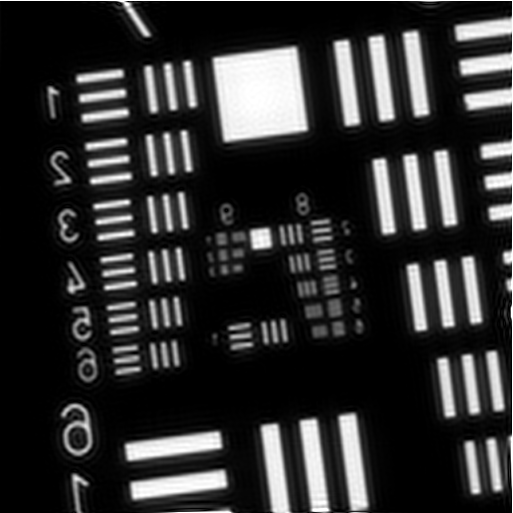}
\caption{ (Top) The central portion of a single unprocessed frame, cropped to $128^2$ pixels.  The processed image with size $1024^2$ pixels is below, revealing significant resolution improvement.}\label{reschart}
\end{figure}

\begin{figure}[htbp]
\centering \includegraphics[width=24pc]{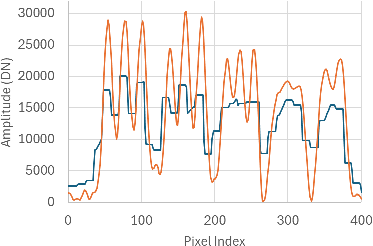} 
\caption{  Profile plots down the horizontal bars of group 8  in Figure~\ref{reschart} showing the difference in resolution between the original and processed images.}\label{profile}
\end{figure}

\section{Application to Resolution Chart}
While resolution improvement is apparent in the previous example, it is difficult to obtain a resolution measurement from the biological sample.  As such, a resolution chart was placed under the microscope and viewed with the same 4X lens.  As before 80 slightly-displaced images were created with diagonal steps of 989.95~nm in an `X' pattern.  The images were cropped, up-sampled by a factor of $M_{SS}=8$, and aligned.  Deconvolution was applied to the combined image.  The central portion of the processed result is shown in Figure~\ref{reschart} (right) alongside the same region of the unprocessed image (left).  Using the bars as a measure, the resolution improves from Group 7, element 6 (228 lp/mm) to Group 8, element 4 (362 lp/mm).   However, as seen in the image, this is a difficult determination when the bars are pixelized.  Figure~\ref{profile} shows profile plots through the horizontal bars in group 8 to demonstrate resolution improvement.

\begin{figure}[htbp]
\centering\includegraphics[width=18pc]{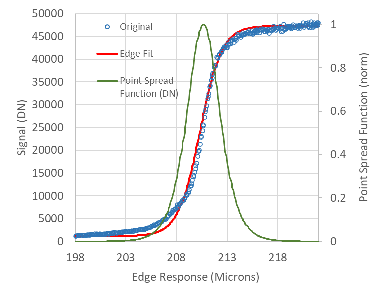}\hspace{2mm} \includegraphics[width=18pc]{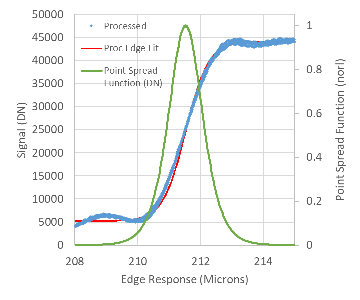}
\caption{ The left plot shows the edge response function, the edge fit and the point spread function for the original image.  The right plot shows the same data for the processed image.  Note the difference in x-axis scale for the two plots. The full-width half maximum values of the PSFs are 3.78 and 1.26 microns respectfully.}\label{edge}
\end{figure}

For improved measurement, we employ the slant edge method.~\cite{aip_2022,laura}  In short, profile plots were taken along a slanted edge in the image and shifted spatially to form an edge-spread function.  A sigmoid function is fitted to the curve.  The edge spread functions are shown in figure~\ref{edge} as well as the fit from the processed image.  The spatial derivative of this curve is a one-dimensional point spread function.  The resolution is then derived from the fitting parameters of the sigmoid function.~\cite{aip_2022}  The resolution of the unprocessed image was determined to be 2.30 pixels or 3.73 microns.  The resolution of the processed image was determined to be 6.57 super-sampled pixels, or 1.39 microns.

The Abbe diffraction limit of the system~\cite{abbe} is found from the equation $  \theta=  \lambda/ 2 NA$ where $NA$ is the numerical aperture of the lens and $\lambda$ is the wavelength of light. Using the central wavelength of 650~nm and NA=0.13, the optical diffraction limit of this lens is 2.5 microns.  The image resolution produced using ImPASS is a factor of 1.79 below this value.  

These values are similar to resolution improvements achieved with Fourier Ptychography  and Structured Illumination Microscopy while having the advantage of simplicity.  ImPASS only requires a translation stage, a common accessory for research facilities and does not require special optics or illumination.  Whereas these methods are fundamentally limited to achieving a factor of two better than the diffraction limit, ~\cite{konda, saxena}  ImPASS subceeded the diffraction limit by a factor of 2.63 in a laboratory test.  Greater resolution can be achieved using methods like STED and SMLM, but with considerably greater cost and complexity.

\section{Comments}
Image Phase Aligned Super-sampling was applied to sets of displaced microscope images taken with a 4X lens.  The processed image of the porcine cornea sample displayed resolved features that were not fully resolved in a truth image captured using the 10X lens.  This suggests that the processing produced a qualitative resolution improvement of 2.5.  To acquire a quantitative measurement of resolution improvement, a USAF resolution  chart was used as the sample.  Using the slant-edge technique, the resolution of the processed image was 1.39 microns, improving the resolution by a factor of 2.68.  This subceeds the diffraction limit of by a factor of 1.79, validating that ImPASS presents as a viable method of super-resolution microscopy.  In contrast to other super-resolution microscopy methods, ImPASS uses passive illumination and a translation stage to create the necessary image set.  This makes the method comparatively inexpensive and easier to implement.  The source of illumination is left for the user to choose.  

\section*{Acknowledgments}
The author greatly appreciates the support from  Dr. Giuliano Scarcelli and Dr. Jiarui Li with the Fischell Department of Bioengineering, University of Maryland, College Park, MD, USA for both creating the image sets used in this study and general advisement.


\end{document}